\documentclass[prl,twocolumn,showpacs,superscriptaddress,a4paper]{revtex4} 
\usepackage{graphicx}
\usepackage{psfrag}
\usepackage{amsmath}
\usepackage{amssymb}
\usepackage{amsfonts}

\topmargin= - 1.0cm

\newcommand{\bb}[1]{{\mathbb #1}}
\newcommand{\mb}[1]{{\mathbf #1}}
\newcommand{\mc}[1]{{\mathcal  #1}}
\renewcommand{\tilde}{\widetilde}

\begin{document}
%
%
\title{Macroscopic current fluctuations in stochastic lattice gases}

\author{L. Bertini}
\affiliation{Dipartimento di Matematica, Universit\`a di Roma
``La Sapienza", Piazza A. Moro 2, 00185 Roma, Italy}

\author{A. De Sole}
\affiliation{Department of Mathematics, Harvard University, 
1 Oxford St., Cambridge, MA 02138, USA 
}

\author{D. Gabrielli}
\affiliation{Dipartimento di Matematica, Universit\`a dell'Aquila,
67100 Coppito, L'Aquila, Italy}

\author{G. Jona--Lasinio}
\affiliation{Dipartimento di Fisica and INFN,
Universit\`a di Roma ``La Sapienza", Piazza A. Moro 2, 00185 Roma,
Italy}

\author{C. Landim}
\affiliation{IMPA, Estrada Dona Castorina 110, J. Botanico, 22460 Rio
de Janeiro, Brazil, \\ and CNRS UMR 6085, Universit\'e de Rouen,
76128 Mont--Saint--Aignan Cedex, France}


%
     %

\begin{abstract}
We study current fluctuations in lattice gases in the macroscopic
limit extending the dynamic approach to density fluctuations developed
in previous articles.  More precisely, we derive large deviation
estimates for the space--time fluctuations of the empirical current
which include the previous results.  
Large time asymptotic estimates for the fluctuations of the time
average of the current, recently established by Bodineau and Derrida,
can be derived in a more general setting. 
There are models where we have to modify their estimates 
and some explicit examples are introduced.
\end{abstract}
%
%


\maketitle

%
%

In a series of recent papers \cite{BDGJL1,BDGJL2,BDGJL3} we have
developed a dynamical approach to macroscopic fluctuations of
thermodynamic variables in non equilibrium steady states of stochastic
lattice gases.  The main object studied in these papers is the
empirical density for which we established a large deviation
principle. From this principle we obtained several results:
we have shown that the entropy functional satisfies a Hamilton--Jacobi
equation, we have extended Onsager--Machlup theory \cite{OMA} and the
minimum dissipation principle \cite{ONS1} to stationary non
equilibrium states. 

Another observable of great physical interest is the macroscopic
current flowing through the system \cite{LLL,DDR2,DDR1,PJSB,PJSB2}.
In the present article we develop, in the same spirit of
\cite{BDGJL1,BDGJL2}, a large deviation principle for the empirical
current and discuss some of its consequences. We discuss the
asymptotic probability, as the number of degrees of freedom increases,
of observing a space--time dependent current profile different from
the typical value.  This new principle includes the previous one for
the density and leads to a unified approach to fluctuations in steady
states both equilibrium and non equilibrium. From this asymptotics we
derive a more general form of a large fluctuation principle recently
obtained by Bodineau and Derrida \cite{bd} for the large time
fluctuations of the time averaged current.
We show that there are models where one has to replace the
large deviation functional obtained in \cite{bd}, that we denote by
$U(J)$, by its convex envelope $U^{**}(J)$. 
We discuss some examples where $U^{**}< U$; this is naturally
interpreted as a dynamical phase transition. However more complex
situations are also conceivable.

In a forthcoming more detailed paper \cite{BDGJL4} we shall study also
the behavior of the current large deviation functional under time
reversal and connect it to the well known fluctuation theorem for
entropy production of Gallavotti and Cohen \cite{gc,k,LS}.
 
As the basic microscopic model we consider a stochastic lattice gas
with a weak external field and particle reservoirs at the boundary.
More precisely, let $\Lambda\subset\bb R^d$ be a smooth domain and set
$\Lambda_N = N\Lambda\cap \bb Z^d$; we consider a Markov process on
the state space $X^{\Lambda_N}$, where $X$ is a subset of $\bb N$,
e.g.\ $X=\{0,1\}$ when an exclusion principle is imposed.  The number
of particles at the site $x\in\Lambda_N$ is denoted by $\eta_x \in X$
and the whole configuration by $\eta\in X^{\Lambda_N}$.  The dynamics
of the particles is described by a continuous time Markov process on
the state space $X^{\Lambda_N}$ with transition rates $c_{x,y}(\eta)$
from a configuration $\eta$ to $\sigma^{x,y} \eta$ that is the
configuration obtained from $\eta$ by moving a particle from $x$ to a
nearest neighbor site $y$. Similar rates $c^\pm_{x}$ describe the
appearance or loss of a particle at the boundary site $x$. The
reservoirs are characterized by a chemical potential $\gamma$. We
assume that the rates satisfy the local detailed balance condition
\cite{LS} with respect to a Gibbs measure associated to some
Hamiltonian $\mc H$.  Typically, for a non equilibrium model, we can
consider $\Lambda$ the cube of side one and the system under a
constant force $E/N$.  Moreover we choose the chemical potential
$\gamma$ so that $\gamma(y/N) = \gamma_0$ if the first coordinate of
$y$ is $0$, $\gamma(y/N) = \gamma_1$ if the first coordinate of $y$ is
$N$, and impose periodic boundary conditions in the other directions
of $\Lambda$.

The fluctuation theory of stochastic lattice gases, as discussed below,
is expected to apply to a wider class of non equilibrium systems with
conservation laws.  
For instance, the Hamiltonian system discussed in \cite{epr} is reduced
to the analysis of a Markov process. 
We introduce the empirical density $\pi^N$ associated to a microscopic
configuration $\eta\in X^{\Lambda_N}$ by requiring for each smooth
function $G:\Lambda \to  \bb R$,
$$
\langle \pi^N, G \rangle = \int_{\Lambda} \!du\: \pi^N(u) \, G(u) =
\frac{1}{N^{d}} \sum_{x\in\Lambda_N} G(x/N) \eta_x 
$$
so that $\pi^N(u)$ is the local density at the macroscopic point
$u=x/N$ in $\Lambda$.
 
Consider a sequence of initial configurations $\eta^N$ such that
$\pi^N(\eta^N)$ converges to some density profile $\rho_0$. 
Under diffusive scaling the empirical density at time $t$ converges,
as $N\to\infty$, to $\rho =\rho(t,u)$ which is the solution of
\begin{equation}
\partial_t \rho
= \nabla \cdot \Big[ \frac 12  D(\rho) \nabla \rho
- \chi(\rho) \nabla V  \Big]
= \cal {D} (\rho)
\label{H}
\end{equation}
with initial condition $\rho_0$.  Here $D$ is the diffusion matrix,
given by the Green--Kubo formula, see \cite[II.2.2]{sp}, $\chi$ is the
conductivity, obtained by linear response theory, see
\cite[II.2.5]{sp}, and $\nabla V$ the external field.  
We emphasize that these transport coefficient are defined in
terms of the equilibrium Gibbs measure. 
In particular if we denote by $S_0 (\rho)$ the entropy associated to
$\mc H$, the usual Einstein relation $D(\rho)=R^{-1}(\rho)\chi(\rho)$
holds, here $R(\rho) = S_0''(\rho)^{-1}$ is the compressibility.  
The interaction with the reservoirs appears as a
boundary condition to be imposed on solutions of (\ref{H}). More
precisely, we require that $S_0'(\rho(u))=\gamma(u)$,
$u\in\partial\Lambda$, here $\partial\Lambda$ denotes the boundary of
$\Lambda$ and we recall that $\gamma$ is the chemical potential
of the reservoirs. 
The non equilibrium stationary profile $\bar\rho$ is the unique stationary
solution of (\ref{H}). 

Denote by $\mu^N$ the non equilibrium stationary ensemble,
i.e.\ the invariant measure of the microscopic process.
The probability of density fluctuations can be described, for large
$N$, by an entropy functional $S(\rho)$ which depends on
the local density of particles $\rho$.  
The stationary profile $\bar\rho$
corresponds to a critical point of $S$ so that it is a minimum of the
entropy; here we use the probabilist sign convention for $S$ which is
opposite to physicist convention.   Hereafter, we
normalize the entropy $S(\rho)$ so that $S(\bar\rho)=0$.
As shown in \cite{BDGJL1,BDGJL2},
the entropy functional $S$ satisfies the following Ha\-mil\-ton--Jacobi
functional derivative equation
\begin{equation}
\label{HJ}
\frac 12 \Big \langle \nabla \frac {\delta S}{\delta \rho} ,
\chi (\rho) \nabla \frac {\delta S}{\delta \rho} \Big \rangle
+ \Big \langle \frac {\delta S}{\delta \rho} ,
{\cal {D} (\rho)}\Big\rangle = 0
\end{equation}
where $\langle \cdot,\cdot\rangle$ means integration with respect to
the space variable $u\in\Lambda$. It shows that
once we know the hydrodynamic equation, we can obtain a closed
macroscopic description.

The probability to observe a macroscopic trajectory different from the
hydrodynamic behavior (\ref{H}) is exponentially small in $N^d$ and
given by
\begin{equation}
\label{LD}
\bb P_\mu^N \big( \pi^N \approx \rho, \; t\in [0,T] \big)
\sim \exp\big\{- N^d I_{[0,T]} (\rho) \big\}
\end{equation}
where $\approx$ denotes closeness in some metric, $\sim$ logarithmic
equivalence as $N\to\infty$, and
$\bb P_\mu^N$ stands for the distribution of the stationary
process, namely we consider an initial condition distributed according
to the invariant measure $\mu^N$. 
The rate functional $I_{[0,T]}(\rho)$ is given by
\begin{equation}
\label{f3}
I_{[0,T]}(\rho)\;=\;
S(\rho(0)) + \frac 12 \int_{0}^T \!dt\:
\langle \nabla H, \chi(\rho) \nabla H \rangle  \;\:
\end{equation}
where the external potential $H$ has to chosen so that $\rho$ solves
\begin{equation}
\label{f2}
\partial_t \rho
= \mc D (\rho) -\nabla \cdot \big( \chi(\rho)  \nabla H \big)
\end{equation}
with the same boundary conditions as the hydrodynamic equation
(\ref{H}). Equations (\ref{LD})--(\ref{f2}) represent a
dynamical generalization of Einstein formula for thermodynamic
fluctuations.

We give now a parallel discussion for the current.
Denote by $\mc N^{x,y}_t$ the number of particles that jumped from $x$ to
$y$ in the macroscopic time interval $[0,t]$. Here we adopt the convention that
$\mc N^{x,y}_t$ represents the number of particles created at $y$ due to
the reservoir at $x$ if $x\not\in \Lambda_N$, $y\in\Lambda_N$ and that
$\mc N^{x,y}_t$ represents the number of particles that left the system at
$x$ by jumping to $y$ if $x\in \Lambda_N$, $y\not \in\Lambda_N$. The
difference $J^{x,y}_t = \mc N^{x,y}_t - \mc N^{y,x}_t$ represents the total
current across the bond $\{x,y\}$ in the time interval $[0,t]$. In
other words, given a path $\eta(s)$, $0\le s\le t$, the instantaneous 
current $d J^{x,y}_{t}/ dt $ is a sum of $\delta$--function localized
at the jump times across the bond $\{x,y\}$ with weight $+1$,
respectively $-1$, if a particle jumped from $x$ to $y$, respectively
from $y$ to $x$.  

Fix a macroscopic time $T$ and denote by $\mc J^N$ the empirical measure
on $[0,T]\times \Lambda$ associated to the current. For smooth
vector fields $G=(G_1, \dots, G_d)$, the integral of $G$ with respect to
$\mc J^N$, denoted by $\mc J^N (G)$, is given by
\begin{eqnarray*}
\mc J^N (G) &=& \int_0^T dt \int _{\Lambda} du \, G(t,u) \cdot \mc J^N(t,u)  \\
&=& \frac{1}{N^{d+1}} \sum_{i=1}^d  \sum_{x} \int_0^T\!dt\:  G_i(t, x/N) \,
{\frac{d J_t}{dt}}^{x,x+e_i}
\end{eqnarray*}
where $\cdot$ is the inner product in $\bb R^d$, $e_i$ is the canonical
basis, and we sum over all
$x$ such that either $x\in \Lambda_N$ or $x+e_i\in \Lambda_N$. 
We normalized $\mc J^N$ so that it is  finite as $N\to\infty$. 

Given a density profile $\rho$ let us denote by 
\begin{equation}
\label{f8}
J(\rho) = -\frac 12  D(\rho) \nabla \rho  + \chi(\rho) \nabla V
\end{equation}
the current associated to $\rho$. The hydrodynamic equation (\ref{H})
can then be written as $\partial_t \rho + \nabla \cdot J(\rho)=0$.
If we consider an initial configuration $\eta^N$
such that the empirical density $\pi^N(\eta^N)$ converges to some
density profile $\rho_0$ and denote by $\rho(t)$ the solution of
(\ref{H}), then the empirical current $\mc J^N(t)$
converges, as $N\to \infty$, to $J(\rho(t))$, the current associated
to the solution of the hydrodynamic equation (\ref{H}). If
we let the macroscopic time diverge, $t\to \infty$, we have 
$J(\rho(t)) \to J(\bar\rho)$.

We next discuss the large deviation properties of the empirical
current.  Fix a smooth vector field $j:[0,T]\times \Lambda \to \bb
R^d$ and a sequence of configurations $\eta^N$ whose empirical density
converges to some profile $\rho_0$. The large deviation principle for
the current states that
\begin{equation}
\label{f1}
\bb P_{\eta^N}^N \big( \mc J^N (t,u) \approx j (t,u) 
\big) 
\sim \exp\big\{ - N^d \, \mc I_{[0,T]}(j)\big\}
\end{equation}
where the rate function is given by 
\begin{equation}
\label{Ic}
\mc I_{[0,T]}(j)\;=\; \frac 12 \int_0^T \!dt \,
\big\langle [ j - J(\rho)  ], \chi(\rho)^{-1}
[ j - J(\rho) ] \big\rangle
\end{equation}
in which $J(\rho)$ is given by (\ref{f8}) and $\rho=\rho(t,u)$ is
obtained by solving the continuity equation 
$\partial_t\rho +\nabla\cdot j =0$ with initial condition $\rho(0)=\rho_0$. 
Of course there are
compatibility conditions to be satisfied, for instance if we have
chosen a $j$ such that $\rho(t)$ becomes negative for some $t\in
[0,T]$ then $\mc I_{[0,T]}(j) = +\infty$.

We present here a heuristic derivation of (\ref{f1})--(\ref{Ic}).
Fix a current $j$; in order to make $j$ typical, we introduce an 
external field $F$. Let $\rho$ be the solution of
\begin{equation}
\label{f4}
\left\{
\begin{array}{l}
\partial_t \rho + \nabla\cdot j = 0 \\
\rho(0,u) = \rho_0(u)
\end{array}
\right.
\end{equation}
and $F: [0,T]\times \Lambda \to\bb R^d$ be the vector field such that
\begin{eqnarray*}
j &=& J(\rho) + \chi(\rho) F \\
&=& - \frac 12 D(\rho)\nabla \rho
+ \chi(\rho) \{ \nabla V + F \} 
\end{eqnarray*}
We introduce a perturbed measure $\bb P^{N,F}_{\eta^N}$ which is obtained by
modifying the rates as follows
$$
c_{x,y}^F(\eta) \;=\; c_{x,y}(\eta) \, e^{N^{-1} F(t, x/N) \cdot (y-x)}
$$
Following a similar argument as the one for the large deviation principle of
the empirical density \cite{BDGJL2}, one can show that
\begin{eqnarray*}
\frac{d \bb P_{\eta^N}^N }{d\bb P_{\eta^N}^{N,F}}
& \sim& \exp\Big\{-N^d \frac 12  \int_0^T \!dt \: 
\langle F , \chi(\rho) F \rangle \Big\} \\
&=& \exp\big\{ - N^d \mc I_{[0,T]} (j) \big\}
\end{eqnarray*}
Moreover, under $\bb P_{\eta^N}^{N,F}$, as $N\to\infty$, $\mc J^N$
converges to $j$. Therefore,
\begin{eqnarray*}
\!\!\!\!\!\!\!\!\!\!\! &&
\bb P_{\eta^N}^N \Big( \mc J^N (t,u) \approx j (t,u), 
\;\; (t,u) \in [0,T]\times \Lambda \Big) \\
\!\!\!\!\!\!\!\!\!\!\! && \quad
= \bb P^{N,F}_{\eta^N} 
\Big( \frac {d\bb P_{\eta^N}^N}{d\bb P_{\eta^N}^{N,F}}
\; \mb 1\{\mc J^N \approx j\} \Big) \sim e^{-N^d \mc I_{[0,T]} (j)}
\end{eqnarray*}

We emphasize that now we allow non--gradient external
fields $F$, while in the large deviation principle for the empirical
density (\ref{LD}), it is sufficient to consider gradient external fields
\cite{KOV,kl}. The latter is therefore a special case and can be
recovered from (\ref{f1})--(\ref{Ic}).

We want to study the fluctuations of the time average of the empirical
current over a large time interval $[0,T]$; the corresponding
probability can be obtained from the space--time large deviation
principle (\ref{f1}). 
Fix some divergence free vector field $J=J(u)$ constant in time and 
denote by $\mc A_{T,J}$ the 
set of all currents $j$ such that $T^{-1} \int_0^T \!dt \: j(t,u) = J(u)$. 
The condition of vanishing divergence on $J$ is required by the local
conservation of the number of particles.
By the large deviations principle (\ref{f1}), for $T$
large  we have
\begin{eqnarray}
\label{LT}
\bb P_{\eta^N}^N \Big( \frac 1T \int_0^T \!dt \; \mc J^N (t) \approx J 
\Big) \sim  \exp \big\{-N^d T \Phi (J) \big\}
\end{eqnarray}
where the logarithmic equivalence is understood by sending {\em
first\/}  $N\to\infty$ and {\em then\/} $T\to \infty$. In
\cite{BDGJL4} we show that for the so--called 
zero range process the limits can be taken in the opposite order; we
expect this to be true in general. The functional $\Phi$ is given by
\begin{equation}
\label{limT}
\Phi (J)
 = \lim_{T\to\infty} \; \inf_{j\in \mc A_{T,J}} 
\frac 1T \; \mc I_{[0,T]} (j)
\end{equation}
By a standard sub--additivity argument it is indeed easy to show the
limit exists.
We now prove that $\Phi$ is a convex functional.
Let $J=pJ_1+(1-p)J_2$: we want to show that
$\Phi(J)\leq p\Phi(J_1)+(1-p)\Phi(J_2)$.
Let us call $(j_1(t),\rho_1(t)),\ t\in[0,pT]$ 
(respectively $(j_2(t),\rho_2(t)),\ t\in[0,(1-p)T]$)
the optimal path of current and density which implements the $\inf$
in (\ref{limT}) at $J_1$ (resp. $J_2$).
We then consider a path $(j(t),\rho(t))$ which spends a time interval $pT$
following $(j_1,\rho_1)$ and a time interval $(1-p)T$ 
following $(j_2,\rho_2)$ (and a finite time to go continuously from 
$(j_1,\rho_1)$ to $(j_2,\rho_2)$).
With such a path we get
\begin{eqnarray*}
\Phi(J) &\leq& \frac{1}{T} \mc{I}_{[0,pT]}(j_1)
+\frac{1}{T} \mc{I}_{[0.(1-p)T]}(j_2) \\
&\leq& p\Phi(J_1) + (1-p)\Phi(J_2) + \epsilon
\end{eqnarray*}
In the last step we used the existence of the limit (\ref{limT}).

We next study the variational problem on the right hand side of 
(\ref{limT}). We begin by deriving an upper bound. 
Given $\rho=\rho(u)$ and  $J=J(u)$, $\nabla\cdot J=0$,
let us introduce the functionals
\begin{eqnarray}
\label{cU}
\mc U (\rho,J) &=& \frac12 \langle J - J(\rho), \chi(\rho)^{-1} 
[J - J(\rho)] \rangle
\\
\label{U}
U(J)& = &\inf_\rho \; \mc U(\rho, J)
\end{eqnarray}
where the minimum in (\ref{U}) is carried over all profiles $\rho$
satisfying the boundary conditions and $J(\rho)$ is given by (\ref{f8}).
When $J$ is constant, that is in the one--dimensional case, 
the functional $U$ is the one introduced in \cite{bd}.

By choosing a suitable path $j(t,u)\in \mc A_{T,J}$ we first show that 
\begin{equation}
\label{1ub}
\Phi(J) \le U(J)
\end{equation}
The strategy is quite simple.  
Let $\hat\rho=\hat\rho(J)$ be the density profile which minimizes the
variational problem (\ref{U}). 
Given the initial density profile $\rho_0$, we construct
a path $j=j(t,u)$, $(t,u)\in [0,T]\times\Lambda$ as follows
\begin{equation*}
j(t) = \left\{
\begin{array}{ccl}
\hat \jmath &\textrm{if} & 0\le t < \tau \\
\vphantom{\Big\{}
\frac{T}{T -2 \tau} \: J &\textrm{if} & \tau \le t < T -\tau \\
- \hat \jmath &\textrm{if} & T -\tau  \le t \le T 
\end{array}
\right.
\end{equation*}
where $\hat\jmath$ is a vector field such that $\tau \nabla \cdot\hat\jmath =
\rho_0 -\hat\rho$ and $\tau>0$ is some fixed time.
It is now straightforward to verify that $j\in \mc A_{T, J}$, 
as well as $\lim_{T\to\infty} \frac{1}{T} \: \mc I_{[0,T]}(j) = U(J)$.

From (\ref{1ub}) and the convexity of $\Phi(J)$ it immediately follows
that
\begin{equation}
\label{ub}
\Phi(J) \le U^{**}(J)
\end{equation}

We next discuss a lower bound for the variational problem (\ref{limT}). 
We denote by ${\tilde {\mc U}}$ and  ${\tilde U}$ the same functionals
as in (\ref{cU})--(\ref{U}), but now defined on the space of all
currents without the conditions of vanishing divergence. 
Let also ${\tilde U}^{**}$ be the convex envelope of $\tilde U$. 

Let  $j\in\mc A_{T,J}$, by the convexity of ${\tilde {U}}^{**}$
in the set of all currents, we get
\begin{eqnarray}
\label{lb}
\frac 1T \mc I_{[0,T]}(j)  
&= & 
\frac 1T \int_0^T\!dt  \: {\tilde {\mc U}} (\rho(t),j(t))
\ge 
\frac 1T \int_0^T\!dt  \: {\tilde U} (j(t))
\nonumber
\\
&\ge &
\frac 1T \int_0^T\!dt  \: {\tilde U}^{**} (j(t))
\ge {\tilde U}^{**}(J)
\end{eqnarray}

The upper and lower bounds (\ref{ub}) and  (\ref{lb}) are, in general,
different. For a divergence free $J$ we have ${\tilde U}(J)= {U}(J)$
but since the convex envelopes are considered in different spaces, 
we only have ${\tilde U}^{**}(J) \le {U}^{**}(J)$. 

To understand the physical meaning of the convex envelope in
(\ref{ub}), suppose $J= p J_1 + (1-p)J_2$ 
and $U(J) > U^{**} (J)= p U(J_1) + (1-p) U(J_2)$ for some
$p,J_1,J_2$.  The values $p,J_1,J_2$ are determined by $J$ and $U$.
In addition we assume that $U^{**} = {\tilde U}^{**}$.  If we
condition on observing an average current $J$, the corresponding
density profile is not determined, but rather we observe with
probability $p$ the profile $\hat\rho(J_1)$ and with probability $1-p$
the profile $\hat\rho(J_2)$.  
When $U$ is not convex we have thus a situation in which the
time averaged current $J$ is realized with the 
coexistence of two dynamical regimes: we have a dynamical phase transition.

The derivation of the upper bound shows that, if $U$ is not convex,
our result differs from the one in \cite{bd}.
On the other hand if ${\tilde U}^{**}(J) < {U}^{**}(J)$ it
is possible that one can improve it by exploring currents with non
vanishing divergence.  In such a situation it is not clear to us if
$\Phi$ can be directly related to $U$.

We can consider the large time behavior of the empirical current as
in equation (\ref{LT}) with the additional constraint 
that the associated density is asymptotically time independent. 
In such a case it is not difficult to show that (\ref{LT}) holds with 
$\Phi=U$. With this extra constraint we are in fact forbidding
the system from oscillating.

In the models where the diffusion coefficient $D(\rho)$ is constant
and the mobility $\chi(\rho)$ is concave, for example in the symmetric
simple exclusion where $\chi=\rho(1-\rho)$, it is not difficult to see
that ${\tilde U}$ is convex. Therefore in these cases $\Phi=U$.  
In \cite{BDGJL4} we shall show that in the 
Kipnis-Marchioro-Presutti model \cite{KMP,BGL} at equilibrium 
$U(J)$ is convex while $\tilde{U}(J)$ is not.

We next discuss an example, with a non--concave $\chi$, 
where the functional $U$ is not convex.
We fix equilibrium boundary conditions $\rho_0=\rho_1=\bar{\rho}$. 
We take
$D(\rho)=1$ and $\chi(\rho)$ a smooth function with
$\chi(0)=\chi(1)=0$ which has a non concave part 
$\chi(\rho)=K e^{-  C \rho}$, for $\rho$ in a given interval,
where $C$ is a positive parameter.
An explicit calculation gives that
\begin{equation} 
U(J)\ =\ \frac{1}{C^2}\, e^{ C\bar{\rho} } F(C J )
\end{equation}
where $F(z)=z \frac{e^z-1} {e^z+1}$, if $J$ is in an appropriate interval.
Since the second derivative of $F(z)$
can be negative, $U(J)$ is not convex.

It is a pleasure to thank T.\ Bodineau, B.\ Derrida, G.\ Gallavotti,
E.\ Presutti, C.\ Toninelli, and S.R.S.\ Varadhan for useful discussions. 
The authors acknowledge the support of COFIN MIUR 2002027798.

\end{document}